\documentstyle[12pt]{article}
\begin{document}
\begin{titlepage}
\title{Scalar Perturbations in a String Inspired Inflationary
Scenario}
\author{Carlos
Eduardo Magalh\~aes Batista\thanks{e-mail: batista@ifi.unicamp.br}\\
\mbox{\small Instituto Gleb Wathagin}\\
\mbox{\small Universidade Estadual de Campinas}\\
\mbox{\small Campinas - S\~ao Paulo}\\
\mbox{\small Brazil}\\
\mbox{\small and}\\
J\'ulio C\'esar Fabris\thanks{e-mail: fabris@cce.ufes.br}\\
\mbox{\small Departamento de F\'{\i}sica}\\
\mbox{\small Universidade Federal do Esp\'{\i}rito Santo}\\
\mbox{\small Vit\'oria - Esp\'{\i}rito Santo - CEP29060-900}\\
\mbox{\small Brazil}}
\maketitle
\begin{abstract}
We consider an inflationary model inspired in the low energy limit
of string theory. In this model, the
scale factor grows exponentially with time. A perturbation study is 
performed,
and we show that there is a mode which displays an exponential growth in
the perturbation of the scalar field.
\end{abstract}
Pacs:98.80
\end{titlepage}
\section{Introduction}
One of most outstanding problems today in Cosmology concerns the
formation of the structures observed in the Universe\cite{r1}. The 
traditional
mechanism for the formations of such structures, the growing of
small fluctuations in the density distribution, can not account for
the existence of galaxies and clusters of galaxies in the context of
the Classical Standard Model. It is now believed that the Inflationary
Scenario can lead to some answer to this question\cite{r2}. In fact, in 
order to
explain the structure formation, we must have fluctuations in the density 
distribution of the order of $10^{-5}$ already in the
beginning of the radiative era. But the initial statistical fluctuations 
must be much less than that. So, in the problem of structure formation it 
is necessary to have a gravitational instability
mechanism
in the primordial Universe in order to grow the perturbations to the desired
value.
\par
So, it seems that
to solve the problem of the structure formation, we must have
a fast growing of the perturbations during some phase before the
radiation dominated period. In this work we investigate such a problem
in the context of an inflationary scenario inspired in a low energy
effective model comming from string theory. In this case, we have
a scalar field $\Psi$ that couples non-minimaly to gravity. We add
a potential term. However, in order to have an exponential growing solution
for the scale factor of the Universe, we include a cosmological
constant. In order to
have some generality in our expressions, we introduce this cosmological
constant as perfect fluid with an equation of state $p = \alpha\rho$,
putting in the end $\alpha = - 1$.
\par
So, our Lagrangian is
\begin{equation}
\label{l1}
{\it L} = \sqrt{-g}e^\Psi\biggr(R + \Psi^{;\rho}\Psi_{;\rho} -
V(\Psi)\biggl) + {\it L_m} \quad .
\end{equation}
Models with a similar field structure has been considered in the
literature\cite{r2'}.
We will show that we can obtain an exponential behaviour for the
scale factor. Our interest, however, is the evolution of the
perturbed quantities around a classical background, which will be
given by the Robertson-Walker metric. Since, we use an
equation of state for the
false vacuum, the perturbations in $\rho$ are zero\cite{r3}. But we can
find non-zero perturbations for the scalar field $\Psi$. The
fluctuations in this scalar field (which is a dilaton field) will be 
responsable for the
perturbations in matter when the inflationary phase ceases and
there is the transition to the radiative phase\cite{r4}.
We can obtain exponential growing modes
for the perturbations in the scalar field during the inflationary phase.
So, this model may contain the desired mechanism for the generation
of the seeds of the structure formation. 
\par
We organize this paper as follows. In section II, we stablish the
background model and its solution; in section III, we perform a
perturbative analysis and determine analytical solutions for the
fluctuations in the physical quantities. We end with the conclusions.
We work with the traditional
formalism of Lifschitz-Khalatnikov, fixing the synchronous coordinate
condition. This has the advange of simplicity giving the
results directly in terms of the physical quantities.
\section{The Background Model}
We will consider in (\ref{l1}) a potential term of the kind $V(\Psi) = 
m^2\Psi^2$, where
$m$ is a mass parameter. It comes out to be convenient, in view of the 
latter perturbation analysis, to
perform a field transformation of the kind $e^\Psi = \phi$. In this case, 
(\ref{l1})
takes the form,
\begin{equation}
\label{l2}
{\it L} = \sqrt{-g}\biggr(\phi R - 
\omega\frac{\phi^{;\rho}\phi_{;\rho}}{\phi} -
V(\phi)\biggl) + {\it L_m} \quad .
\end{equation}
In this expression, we should have $\omega = - 1$, in order to be in 
agreement with
(\ref{l1}). We work with an arbitrary $\omega$ in such a way to be in 
contact with
the Brans-Dicke theory, which is the prototype of the scalar-tensorial 
theories.
Moreover, $V(\phi) = m^2\phi ln^2\phi$. Again, ${\it L_m}$ is the 
lagrangian density	
associated with matter, which for our purposes will be a perfect fluid 
matter, with
the equation of state $p = - \rho$. But, we will introduce this 
hypothesis just in the
end.
\par
From (\ref{l2}), and defining $\phi W(\phi) =  V(\phi)$, we deduce the 
following field equations:
\begin{eqnarray}
R_{\mu\nu} - \frac{1}{2}g_{\mu\nu}R &=& \frac{8\pi }{\phi}T_{\mu\nu} +
\frac{\omega}{\phi^2}(\phi_{;\mu}\phi_{;\nu} - 
\frac{1}{2}g_{\mu\nu}\phi^{;\rho}\phi_{;\rho}) +
\nonumber \\
& &+ \frac{1}{\phi}(\phi_{;\mu;\nu} - g_{\mu\nu}\Box\phi) -
\frac{g_{\mu\nu}}{2}W{\phi} \quad ; \\
\Box\phi &=& \frac{1}{3 + 2\omega}\biggr(8\pi T + \phi^2W' - \phi 
W\biggl) \quad ; \\
{T^{\mu\nu}}_{;\mu} &=& 0 \quad .
\end{eqnarray}
In these equations, the prime represent derivation with respect to the 
field $\phi$ and $T^{\mu\nu}$ comes from the variation of
${\it L_m}$ with respect to the metric.
We consider the spatially flat Robertson-Walker metric,
\begin{equation}
ds^2 = dt^2 - a^2(t)(dx^2 + dy^2 + dz^2) \quad ,
\end{equation}
and the momentum-energy tensor as
\begin{equation}
T^{\mu\nu} = (\rho + p)u^\mu u^\nu - pg^{\mu\nu} \quad .
\end{equation}
The pressure and energy density are linked by a barotropic
equation of state $p = \alpha\rho$.
The differential linking $a(t)$, $\phi(t)$, $p$ and $\rho$ are:
\begin{eqnarray}
3(\frac{\dot a}{a})^2 &=& \frac{8\pi}{\phi}\rho + 
\frac{\omega}{2}(\frac{\dot\phi}{\phi})^2 -
3\frac{\dot a}{a}\frac{\dot\phi}{\phi} + \frac{W}{2} \quad , \\
\ddot\phi + 3\frac{\dot a}{a}\dot\phi &=& \frac{1}{3 + 2\omega}
\biggr(8\pi(\rho + 3p) + \phi^2W' -
\phi W\biggl) \quad ,\\
\dot\rho + 3\frac{\dot a}{a}(\rho + p) &=& 0 \quad .
\end{eqnarray}
\par
Imposing $p = - \rho$, $\phi = \phi_0 = cte$, we obtain the following 
solutions:
\begin{eqnarray}
a &=& a_0e^{Ht} \quad ;\\
16\pi\rho &=& m^2\phi_0ln\phi_0 \quad ; \\
ln\phi_0 &=& - 1 \pm \sqrt{1 + 12\frac{H^2}{m^2}} \quad .
\end{eqnarray}
These solutions represent an inflationary phase, driven by a cosmological 
constant (that can
be interpreted as a false vacuum state) and a constant scalar field.
This scenario is well in spirit of old inflation, with its
advantage and disadvantage. We will not discuss here, for example, the 
mechanism to
end with this inflationary phase, which has an abundant literature. Our 
interest is its consequence
for the evolution for small perturbations during this period, which is 
believed to be limited in time in a more complete theory.
\section{Perturbative Analysis}
Now, we turn to the problem of the perturbation of physical
quantities in this model. Essentially, we have the perturbations in the 
metric $\delta g_{\mu\nu} = h_{\mu\nu}$,
in the scalar field $\delta\phi$ and in the matter $\delta\rho$.
We employ the Lifschitz-Khalatnikov
formalism mainly for two reasons:	
it is simpler mathematically, and it gives the results directly	in
terms of the perturbations in the matter and field quantities. Since in 
this formalism we fix
a frame, we must be sure that our results concern physical
quantities, and not coordinate artifacts\cite{r4}. This can be done by
studying the residual coordinate freedom as we will see below.
\par
So, we introduce in the field equations the
quantities $g_{\mu\nu} = \stackrel{0}{g}_{\mu\nu} + h_{\mu\nu}$,
$\phi = \stackrel{0}{\phi} + \delta\phi$, $\rho = \stackrel{0}{\rho}
+ \delta\rho$, $u^\mu = \stackrel{0}{u^\mu} + \delta u^\mu$, i.e., the 
variables now are the sum of the
background solution and of a small perturbation around it.
We impose the synchronous coordinate condition $h_{\mu 0} = 0$.
We will consider that these perturbations are adiabatic,  with
$p = \alpha\rho$ and $\delta p = \alpha\delta\rho$.
The procedure to deduce the perturbed equations are quite standard,
and the final results are\cite{r6,r6'}:
\begin{eqnarray}
\label{p1}
\dot\Delta + (1 + \alpha)(\Theta - \frac{\dot h}{2}) &=& 0 \quad , \\
\label{p2}
(1 + \alpha)\dot\Theta + (2 - 3\alpha)(1 + \alpha)\frac{\dot a}{a}\Theta
+ \alpha\frac{\nabla^2\Delta}{a^2} &=& 0 \quad ,\\
\ddot\lambda + \dot\lambda(3\frac{\dot a}{a} + 2\frac{\dot\phi}{\phi}) + 
\frac{\lambda}{3 + 2\omega}\biggr(\frac{32\pi \rho}{\phi} + W^{''}\phi^2
- \frac{q^2}{a^2}(3 + 2\omega)\biggl) \nonumber \\
+ \dot h\frac{\dot\phi}{\phi}
- \frac{32\pi \delta\rho}{\phi(3 + 2\omega)} &=& 0 \quad .
\label{37}
\end{eqnarray}
In these equations, $h = \frac{h_{kk}}{a^2}$, $
\Delta = \frac{\delta\rho}{\rho}$,
$\lambda = \frac{\delta\phi}{\phi}$, $\Theta = {\delta u^i}_{,i}$.
The term $q^2$ is provenient from a plane wave expansion $\lambda \propto
exp(iqx)$ where $x$ represent the comoving distance and $q$ is
the wavenumber of the perturbation.
We will work in the conformal time in order to simplify 
the resolution of the equations, that is $dt = ad\tau$. So,
$a = -\frac{1}{H\tau}$.
\par
Imposing $\alpha = - 1$, we can see that the perturbations in
matter are null.
The equation for $\lambda$ can be written as
\begin{eqnarray}
\lambda'' + \lambda'(2\frac{a'}{a} + 2\frac{\phi'}{\phi})
+ \nonumber \\
+ \frac{\lambda}{3 + 2\omega}\biggr(\frac{32\pi\rho a^2}{\phi} - 
4\frac{dW}{d\phi}\phi a^2 -
\frac{d^2W}{d^2\phi}\phi^2 a^2 + q^2(3 + 2\omega)\biggl) = - 
h'\frac{\phi'}{\phi}  
\end{eqnarray}   
where primes represent now derivatives with respect to conformal 
time.  
Considering now the background solutions, this equation takes
the form:
\begin{equation}
 \lambda'' - 2 \frac{\lambda'}{\tau} +  \lambda\biggr(\frac{4}{1+ 2\omega}(
\frac{9 + A/H^2}{\tau^2}) + q^2\biggl) = 0 \quad .
\end{equation}
where $A$ is:
\begin{equation}
 A = \frac{- (1 + 4\omega)\frac{dW}{d\phi}\phi + 3W }{2(3+ 2\omega)} + 
\frac{3W}{2} 
- \frac{d^2W}{d^2\phi}\frac{\phi^2 ( 1 + 2\omega)}{4(3 + 2\omega)}  \quad.
\label{45}
\end{equation} 
Writing $\lambda = \tau^p\gamma$,
where $p=\frac{3}{2}$, is easy to show that our equation becomes:
\begin{equation}
\label{fe}
 \gamma'' + \frac{\gamma'}{\tau} + [(-9/4 + \frac{4(9 + A/H^2)}{
1 + 2\omega})\frac{1}{\tau^2}  + q^2]\gamma = 0 \quad .
\end{equation}
 \par
The equation (\ref{fe}) has solutions under the form of Bessel's
functions. The final result for $\lambda$ is,
\begin{equation}
\lambda = \tau^p\biggr(J_\nu(q\tau) + J_{-\nu}(q\tau)\biggl) \quad .
\end{equation}
where
\begin{equation}
\nu^2 = -\frac{9}{4} + 4(\frac{9 + \frac{A}{H^2}}{1+2\omega}) \quad .
\end{equation}
This solution contains just physical modes. In fact, employing the
synchronous coordinate condition, we have yet a residual coordinate
freedom, given by a coordinate transformation $x^\mu \rightarrow
x^\mu + \chi^\mu$, which preserves the synchronous condition. But,
it can be verified that we can not eliminate the above solutions
using this coordinate freedom.
\par
In some cases, $\nu$ can be purely imaginary, namely, for
$-1.5 < \omega < -1.4$ or $-0.5 < \omega < 31.4$. For the cases where
$\omega$ is $-1.5$ or $-0.5$, the order of the Bessel function goes
to infinity; in the two orther limits, it goes to zero.
\par
We can have a better insight on the nature of these solutions
by investigating their asymptotical behaviour. We consider first
the case where the order of the Bessel function is real.
In the case of small values of $\tau$ (which implies large values
of the cosmic time $t$), we have,
\begin{equation}
\lambda \propto \tau^{\frac{3}{2} \pm \nu} \propto
e^{-(\frac{3}{2} \pm \nu)Ht} \quad .
\end{equation}
We can have an exponential growing for the perturbations if
$\frac{3}{2} \pm \nu < 0$. In the case where $\frac{H^2}{m^2} >> 1$ ,
which represents the condition for having enough inflation\cite{r7},
this means $\omega < - \frac{5}{4}$.
\par
In the case $\tau >> 1$, which means $t << 1$, we have the asymptotical limit
\begin{equation}
\lambda \propto \tau (c_1Cos\tau + c_2Sin\tau) \propto
e^{-Ht}(c_1Cos(e^{-Ht}) + c_2Sin(e^{-Ht})) \quad.
\end{equation}
So, the solutions present an exponential decreasing oscillations
in the beginning, and they can evolve to exponential increasing
modes, depending on the value of $\omega$.
\par
For the case, where the order of the Bessel function is purely imaginary,
we have an oscillatory behaviour in both asymptotical limits.
\section{Conclusions}
The Inflationary Scenario we have described here can be obtained by
considering a dilaton field plus a false vacuum energy density.
This false vacuum is formaly represented by a perfect fluid with
an equation of state $p = - \rho$. In one sense, this field structure
was inspired in the low energy limit of string theory, for which we
have add a potential for the inflaton, and the term representing the
false vacuum. We have obtained an exponential growth for the
scale factor, following closely the spirit of old inflation.
\par
In general, in the de Sitter phase,
the density perturbations is zero\cite{r8}; this result is also true in the
case of the traditional Brans-Dicke theory\cite{r15}. Our model is 
somehow similar to the Brans-Dicke case, since the gravitational coupling 
can be a varying function of time. In this case, even
if the matter perturbations are zero during the de Sitter phase, 
inhomogeneities
in the scalar field, which is linked with the gravitational coupling,
can induce matter inhomogeneities in the end of the inflationary phase.
\par
We have found many different situations for the behaviour of the
perturbation of the scalar field. In some cases, in both asymptotical
limits, this perturbation has oscillatory behaviour, with constant or
decreasing amplitude.
On the other hand, we verified that for values of the parameter $\omega$
lesser than $-\frac{5}{4}$, when $\frac{H^2}{m^2} >> 1$, we can have an 
exponential growth for the perturbation
of the scalar field.
\par
It has been determined recently some cases where this exponential growth
can also occur, but directly for the density contrast of ordinary matter
\cite{r16,r17}. In our case, even if this exponential growth does not
concern directly ordinary matter, the inhomogeneities in the scalar field can
have important consequences for the formation of structures in a latter
phase, after the end of the inflationary regime.
\newline
{\bf Acknowledgements}: We thank Jos\'e Pl\'{\i}nio Baptista and Bert Schroer
for their comments and suggestions. This work was partially supported by
CNPq (Brazil).

\end{document}